# Fourth time's a *XARM*

Arguably, no mission changed X-ray astronomy in as short a time as did *Hitomi*. The planned *X-ray Astronomy Recovery Mission, XARM*, will carry its legacy forward.

**Poshak Gandhi**
*(University of Southampton, Department of Physics & Astronomy, Highfield, Southampton, SO17 1BJ, UK)*

X-ray telescopes are sensitive to cosmic photons thousands of times more energetic than those perceptible to the human eye. Such energetic photons originate in dense and high temperature plasmas found around black holes, in clusters of galaxies, the solar corona, and in planetary aurorae. Million-degree gas in the intergalactic medium accounts for about half of the cosmic baryon budget[1]. X-ray astronomy is our gateway to studying this ubiquitous, high-energy component of the universe.

The Earth's atmosphere is opaque to X-rays, so X-ray telescopes must be placed in space. The X-ray band is broad, spanning a range of ~1,000 in photon energies (~0.3—500 kilo electron Volts [keV]). The commensurate wavelengths are miniscule (~40—0.02 Angstroms) and impose important technical challenges for detecting X-rays and splitting them into their constituent energies. X-ray telescopes lag far behind optical telescopes in terms of their splitting power (or 'resolution' denoted as $R$). The typical spectral resolution $R(=E/\Delta E)$ of charge-coupled device (CCD) detectors in most orbiting X-ray satellites is only about 50 or less. Gratings can improve upon this by dispersing an incoming beam of X-rays, but are most effective at the lowest energies ($E \leq 2$ keV). Above this, the high-energy grating on board NASA's Chandra observatory has achieved a best spectral resolution of $R\sim160$ for important transition energies of iron between 6 and 7 keV. Finer splitting becomes technically challenging at short wavelengths.

Low $R$ values limit our capability to discern the physics and composition of cosmic plasmas. For instance[2]: mapping the dynamics and geometry of material accreting from parsec scales on to supermassive black holes requires a velocity resolution of order few hundred km s$^{-1}$. Similar sensitivity in stellar-mass black holes could distinguish between hotly debated wind launching mechanisms. For comparison, *Chandra*'s high-energy gratings can only resolve turbulent motions of ~2,000 km s$^{-1}$ or more. Detailed diagnostics of the warm-hot intergalactic medium and of turbulent plasmas motions all currently suffer from lack of resolution.

Better finesse requires a step change in technology, and microcalorimeters provide the way forward. Microcalorimeters detect minute heat (hence 'caloric') changes associated with absorption of individual X-ray photons, in order to determine incoming photon energies. They thus do away with the need to disperse out an incoming beam, which, in turn, optimises sensitivity. By cooling the instruments down to temperatures of just 1/20$^{th}$ of a degree above absolute zero, thermal noise is minimised and $R$ values of more than 1,000 can be achieved. The Calorimeter Group at the Goddard Space Flight Centre has pioneered development of microcalorimeter payloads[3].

The idea of placing such instruments in orbit has been around for decades, but it is the Japanese who have led this charge. The *Astro-H* mission was launched on 2016 Feb 17 (thereupon christened as *Hitomi*), and was meant to deliver on the promise of this novel technology. It had a planned mission lifetime of 3 years; but lasted only ~38 days. A series of anomalies, including a design flaw in the attitude control, coupled with human error, led to a fatal cascade of failures on Easter weekend resulting in breakup of the Solar paddles and irretrievable loss of communication and control, ending the mission prematurely.[4]

*Astro-H* was one of the largest international collaborative astronomy missions, involving scientists and engineers from Japan, USA, Canada and Europe. It followed two prior unsuccessful attempts to carry out observations with microcalorimeters – *Astro-E* (2000) and *Suzaku* (2005-15). So this failure was a body blow to many in the X-ray astronomy community, especially in Japan. I was resident in Tokyo and participated in the scientific discussions between 2010-13. At the Japan Aerospace Exploration Agency, I saw just how hard the mission specialists worked behind the scenes; my heart aches for the many Japanese colleagues who had spent decades planning for the mission. Several ended up leaving the field or changing research direction completely. I myself had to abandon plans for a major grant application in 2017 and reorient the research focus of my group.

Nevertheless, the mission was far from a complete failure. Thanks to prudent planning, early observations were carried out of a high priority science target – the Perseus cluster of galaxies located at a distance of ~240 million light-years. This is the X-ray brightest galaxy cluster in the sky and comprises thousands of member galaxies as well as 50-million deg C hot gas sloshing about within the cluster's gigantic gravity well. Studying the dynamics and composition of intracluster gas can shed light on the formation of large-scale structures throughout cosmic time.

*Hitomi*'s capability for such studies surpassed any previous X-ray mission. Its microcalorimeter, named Soft X-ray Spectrometer (SXS), achieved a superb energy resolution $\Delta E$~4.7 eV ($R$~1,400), and revealed a series of bright emission features arising from excited ions in the intracluster plasma (Fig. 1)[5]. X-ray observations are often photon starved, but in the *Hitomi* Perseus observation, there are tens of thousands of detected counts in the complex of Iron emission features alone. The exquisite quality of the data is even more remarkable considering that these early observations were carried out at less than full efficiency, with a safety gate valve blocking a significant fraction of incoming photons.

SXS showed the intracluster plasma in Perseus to be astoundingly calm, constraining the turbulent gas velocity to be just 164±10 km s$^{-1}$. Such velocity precision is unprecedented in X-ray astronomy, and the results are a surprise because there is a supermassive black hole at the cluster core pumping enormous energy into the plasma; yet some as yet unknown process is suppressing this energy from inducing turbulence in its surrounding. This finding has wide implications for the physics of energy cycles ('feedback') spawned by black hole growth, and also suggests that clusters can be used as effective cosmological probes[5].

Additional intriguing results followed a few months later from a search for a radiative transition attributed to decaying dark matter particles. An electromagnetic signal from dark matter would be a monumental discovery, but SXS observations ruled out a prior report of a bright associated feature around $E$~3.5 keV in Perseus[6]; the origin of this putative feature continues to be debated.

A further highlight emerged from chemical analysis of the hot cluster gas. Important signatures of iron-peak elements from Type Ia supernovae (Chromium, Manganese, Nickel and Iron), in particular, have been difficult to clearly isolate prior to *Hitomi*. SXS analysis found that the relative proportion of various elements in the cluster gas is nearly identical to that of the Sun, and showed that this was a result of a mixed distribution of Type Ia progenitor masses[7]. The results are unexpected as there is no a priori reason to expect similar chemical evolutionary histories between our solar neighbourhood and galaxies in the Perseus cluster.

This month [April], the *Hitomi* team published a slew of additional follow-up papers in a special issue of the Publications of the Astronomical Society of Japan. These include data from the handful of other sources observed by *Hitomi*: a high mass X-ray binary, two supernova remnants, the Crab, and the accreting active galaxy at the centre of Perseus (the only accreting black hole observed at high spectral resolution). These data provide tantalising evidence of the tremendous scope of possible new discoveries, including the presence of new spectral features and novel insight on plasma velocities in a variety of extreme environments (https://academic.oup.com/pasj/issue/70/SP2).

For such a short-lived mission, *Hitomi*'s scientific legacy is extraordinary. No other general user observatory has produced several high impact publications from its first-light observations. *Hitomi*'s transformational science output was a catalyst for rapid approval of a successor mission, currently named the *X-ray Astronomy Recovery Mission* (*XARM*, pronounced 'charm')[8]. This is a JAXA-led mission overseen by NASA and was approved for a 2020/21 launch subject to restructuring of the project based upon lessons learnt from *Hitomi*. It will be significantly cheaper than *Hitomi*. The microcalorimeter counterpart on *XARM* is christened 'Resolve' and could utilise space-ready hardware spares from SXS. *XARM*'s unique parameter space is illustrated in Fig. 2.

*XARM* is also an important mission for Europe, as it is a pathfinder towards ESA's next large X-ray observatory *ATHENA*, the *Advanced Telescope for High-ENergy Astrophysics*, aimed for 2028. *ATHENA* will carry a next-generation microcalorimeter with a multiplex of transition edge sensors that will double SXS's spectral resolution capability. We can also expect orders of magnitude improvement in sensitivity based upon a larger collecting area and finer imaging resolution[9]. ESA is also participating in *XARM* with contribution of a filter wheel and other components. Here in the UK, a discussion meeting at the Royal Astronomical Society was organised for the X-ray community to better connect to these frontline missions[10]. Deteriorating real term budgets are undermining curiosity-driven science exploitation possibilities, so a cohesive community strategy is the need of the hour. ESA recently had an announcement of opportunity for scientists to participate in *XARM*, following a similar call by NASA. These are very welcome for broadening participation in the mission.

At present, no other space agency has advanced plans to explore the high spectral resolution X-ray frontier. With *ATHENA*'s launch at least a decade in the future, the global X-ray astronomy community has high hopes for science to be delivered by XARM.

**References**

1. Cen, R., Ostriker, J. P. Where are the baryons? *Astrophys. J.* **514**, 1 (1999).

2. Takahashi, T. et al. *ASTRO-H* White Paper – Introduction. **arXiv:1412.2351** (2014).

3. Kilbourne, C. et al. Design, implementation, and performance of the Astro-H SXS calorimeter array and anticoincidence detector. *JATIS* **4(1)**, 011214 (2018).

Submitted 2018 April.

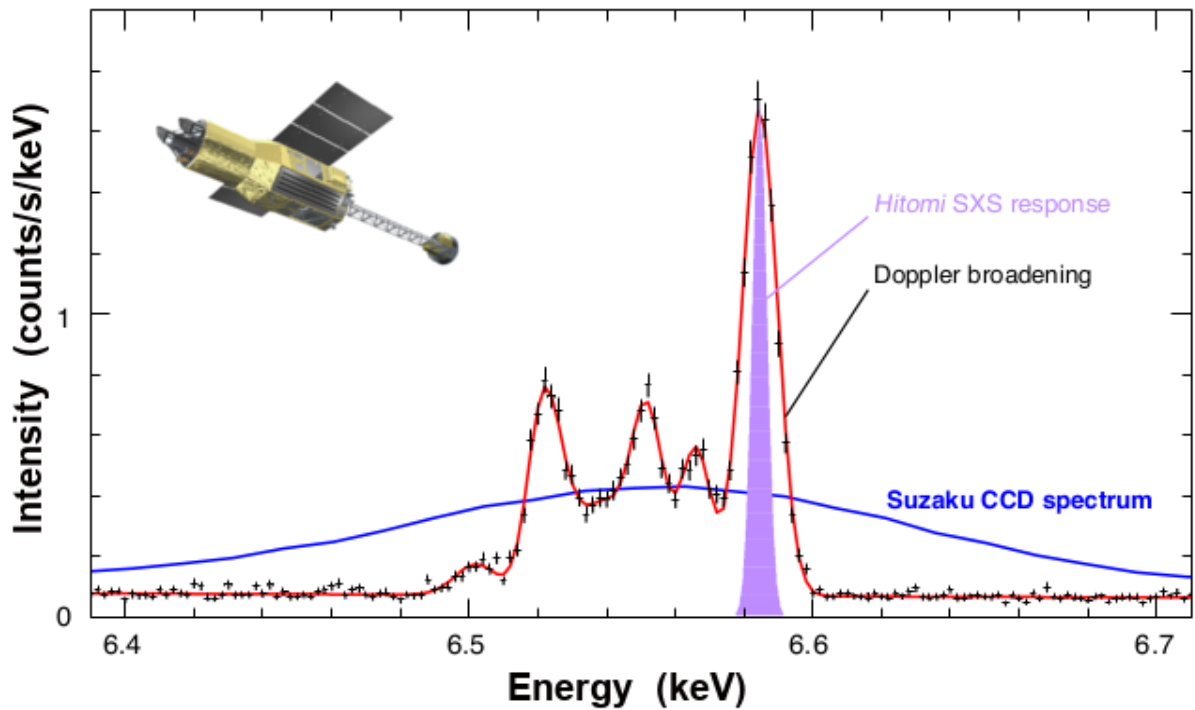

Figure 1: The *Hitomi* spectrum of the Perseus cluster around the strongest emission transitions of ionised Iron (Fe XXV)[5]. *Hitomi*'s microcalorimeter SXS resolved individual transitions and constrained turbulent motions to an unprecedented sensitivity of 10 km s$^{-1}$. The exquisite spectral resolution is denoted in shaded purple, compared to that of current generation CCD detectors in blue (*Suzaku*). *Hitomi* is shown on the top left.

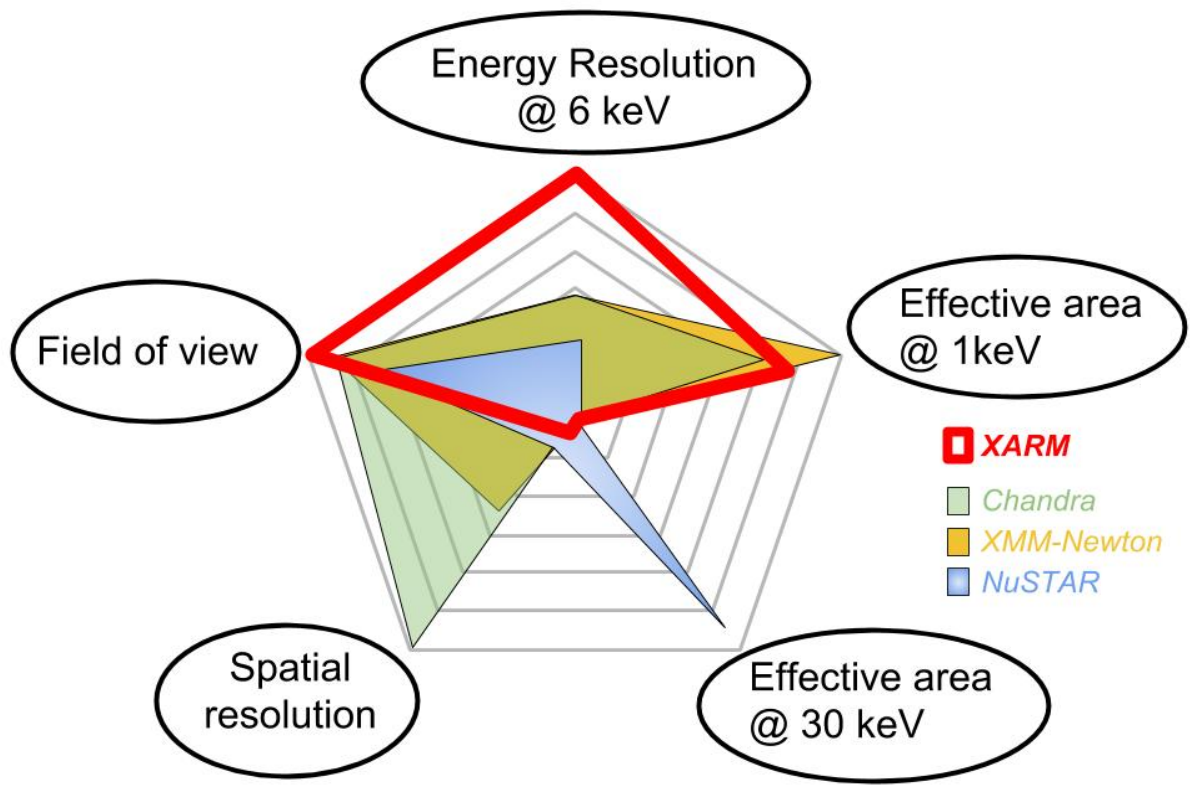

Figure 2: The novel parameter space opened up by the microcalorimeters on *XARM* and *Hitomi*, significantly surpassing all other X-ray missions in terms of spectral resolution above energies of ~2 keV[8].